\documentclass[prl,twocolumn,showpacs]{revtex4}
\usepackage{amssymb}
\usepackage{epstopdf}
\usepackage{graphicx}
\usepackage{dcolumn}
\usepackage{amsmath}
\usepackage{amsfonts}

\begin{document}

\title[Short Title]{Ultrafast dephasing of coherent optical phonons in atomically controlled 
GeTe/Sb$_{2}$Te$_{3}$ superlattices}
\author{Muneaki Hase$^{1,2}$}
\email{mhase@bk.tsukuba.ac.jp}
\author{Yoshinobu Miyamoto$^{1}$}
\author{Junji Tominaga$^{3}$}
\affiliation{$^{1}$Institute of Applied Physics, University of Tsukuba,
1-1-1 Tennodai, Tsukuba 305-8573, Japan}
\affiliation{$^{2}$PRESTO, Japan Science and Technology Agency, 4-1-8 Honcho,
Kawaguchi, Saitama 332-0012, Japan}
\affiliation{$^{3}$Center for Applied Near-field Optics Research, National Institute of 
Advanced Industrial Science and Technology, Tsukuba Central 4, 1-1-1 Higashi, 
Tsukuba 305-8562, Japan}

%\date{\today}

\begin{abstract}
Femtosecond dynamics of coherent optical phonons in GeTe/Sb$_{2}$Te$_{3}$ 
superlattices (SLs), a new class of semiconductor SLs with three different 
states, have been investigated by using a reflection-type pump-probe technique at 
various lattice temperatures.  
The time-resolved transient reflectivity (TR) obtained in as-grown SLs exhibits the 
coherent A$_{1}$ optical modes at 5.10 THz and 3.78 THz, while only 
the single A$_{1}$ mode at 3.68 THz is observed in 
annealed SLs. The decay rate of the A$_{1}$ mode in annealed SLs is strongly 
temperature dependent, while that in as-grown SLs is not temperature dependent. 
This result indicates that the damping of the coherent A$_{1}$ phonons in amorphous 
SLs is governed by the 
phonon-defect (vacancy) scattering rather than the anharmonic phonon-phonon coupling.  
\end{abstract}


\maketitle 
One of the most common materials for optical recording media is 
Ge$_{2}$Sb$_{2}$Te$_{5}$ (GST), in which phase transition between crystalline and 
amorphous phases serve rewritable recording\cite{Yamada91,Yamada00}. Recently, extensive 
theoretical investigation 
on the mechanism of the phase change in GST have been made using molecular dynamics 
simulations \cite{Hegedus08, Sun06, Sun07}. In addition, experimental studies using extended 
x-ray absorption fine structure (XAFS) and Raman scattering measurements have examined  
dynamics of phase transition in GST \cite{Kolobov04, Kolobov06, Baker06, Andrikopoulos07},  
suggesting that the structure of amorphous GST can be described as a cross-section of a 
distorted rocksalt structure with vacancies and the amorphization of GST is due to an umbrella 
flip of Ge atoms from an octahedral position into a tetrahedral one. Moreover, Sun {\it et al.} 
theoretically proposed that the vacancies in the crystalline (cubic) GST are highly ordered and 
layered \cite{Sun06}, followed by the recent prediction of the formation of large voids in the amorphous 
GST films\cite{Sun09}. The experimental information on the existence of vacancies from the lattice dynamical 
point of view, however, has not been explored. 

One of the advantages of GST as the optical recording media is its high speed switching 
of read-write characteristics, whose time scale has been believed to be less than a nanosecond. 
In order to understand and to control the rapid phase change in GST, 
a time-resolved study of phonon dynamics in GST is strongly demanded, however, the 
time-resolved study is still very few \cite{Forst00}. Moreover, 
a new class of semiconductor superlattices (GeTe/Sb$_{2}$Te$_{3}$) with three 
different states have recently been proposed, which will enable us to realize 
reversible transition among the three states by means of the irradiation of laser 
pulses \cite{Chong08}.  

The coherent phonon spectroscopy (CPS) is a powerful tool to study ultrafast dynamics 
of structural phase transitions, occurring within pico- and femtoseconds time scale, 
and in fact, it has been applied to semimetals \cite{Hase02}, ferroelectric 
materials \cite{Hase03,Lu07}, and Mott insulators \cite{Cavalleri04,Kubler07}. 
In the CPS, the pump pulse impulsively generates coherent 
lattice vibration through real or virtual electronic transitions. The pulse length ($\Delta$) 
used should be much shorter than the time period of the lattice vibration, so it is typically 
$\Delta$ $\leq$100 fs to excite phonons with terahertz (THz) frequency. It has been shown 
that the dephasing of the coherent optical phonon is very sensitive to the density of 
vacancy \cite{Hase00}.
Regarding to Ge$_{2}$Sb$_{2}$Te$_{5}$, F\"{o}rst {\it et al}. investigated dynamics of 
phase transition in GST films by using the CPS and found that the appearance of the 
phonon modes was significantly modified upon the structural change among amorphous, 
cubic, and hexagonal structures \cite{Forst00}. In their study, the observed phonon 
modes in GST films were always strongly damped modes, with its dephasing time of 
less than several picoseconds, however, the dynamics of the dephasing of coherent 
phonons in GST compounds have not yet been revealed. 
 
In this paper, we present ultrafast dephasing dynamics of coherent optical phonons 
observed in atomically controlled GeTe/Sb$_{2}$Te$_{3}$ SLs at various 
lattice temperatures. Our motivation for the use of the atomically controlled GeTe/Sb$_{2}$Te$_{3}$ 
SLs is based on the new structural model that Ge$_{2}$Sb$_{2}$Te$_{5}$ is considered as superlattice, which  
consists of two units: one is a Ge$_{2}$Te$_{2}$ layer and 
the other is a Sb$_{2}$Te$_{3}$ layer \cite{Tominaga08,Silva08}. That indicates understanding the role of the 
flip-flop transition of the Ge atom in the distorted simple-cubic unit cell will be promising strategy toward 
the reversible transition by means of the irradiation of ultrashort laser pulses.
Our experiments show that the frequency of the coherent A$_{1}$ optical modes decreases with 
increasing the lattice temperature.  The decay rate (the inverse of the dephasing time) 
of the coherent A$_{1}$ mode increases as the lattice temperature increases in annealed 
(crystalline) SLs, while that in as-grown (amorphous) SLs is almost constant over the wide 
temperature range. Our data demonstrate that randomly distributed vacancies or voids 
indeed exist in the amorphous phase, while they become "ordered" in the crystalline phase. 
The differences of phonon dynamics in GST film and GeTe/Sb$_{2}$Te$_{3}$ SLs are also 
discussed. 

A reflection-type pump-probe measurements using a mode-locked Ti:sapphire laser 
($\Delta$= 20 fs and a central wavelength 850 nm) was employed at the temperature range 
of 5 - 300 K. This system enabled us to detect optical response up to 30 THz bandwidth. 
The average power of the pump and probe beams were fixed at 120 mW and 3 mW, 
respectively, from which we estimated the pump fluence to be 284 $\mu$J/cm$^{2}$ at 120 mW. 
The optical penetration depth of the laser light was estimated to be $\leq$50 nm at 1.46 eV, 
which is longer than the thickness of GST film. This fact indicates that we probe the whole depth of 
the films including the interface to Si substrate, however, the contribution from the Si substrate to 
the signal should be negligibly small because of extremely small optical absorption strength in Si 
at 1.46 eV photon energy\cite{Yu99}. 
The samples used were GST film (18 nm thick) and thin film consisting of superlattice of GeTe and 
Sb$_{2}$Te$_{3}$ layers (GeTe/Sb$_{2}$Te$_{3}$) fabricated using a helicon-wave 
RF magnetron sputtering machine on Si (100) substrate. 
The annealing of the as-grown GeTe/Sb$_{2}$Te$_{3}$ SL films at 503 K (230 $^{\circ}$C) 
for ten minutes changed 
the amorphous states into the crystalline state \cite{Tominaga08}. The TEM measurements 
confirmed that the GeTe/Sb$_{2}$Te$_{3}$ SLs have layered structures with clear interfaces. 
In addition, the structural change of the GeTe/Sb$_{2}$Te$_{3}$ SLs was evident from the 
increase in the reflectivity ($\sim 10 \%$) after the annealing. The excitation of the GST 
and GeTe/Sb$_{2}$Te$_{3}$ SLs  with the 850 nm (= 1.46 eV) laser pulse generates 
photo-carriers across the narrow band gap of $\approx$ 0.5 - 0.7 eV \cite{Lee05}. 
TR signal ($\Delta R/R$) was measured as a function of the time delay $\tau$ after 
excitation pulse. 
\begin{figure}
\includegraphics[width=8.6cm]{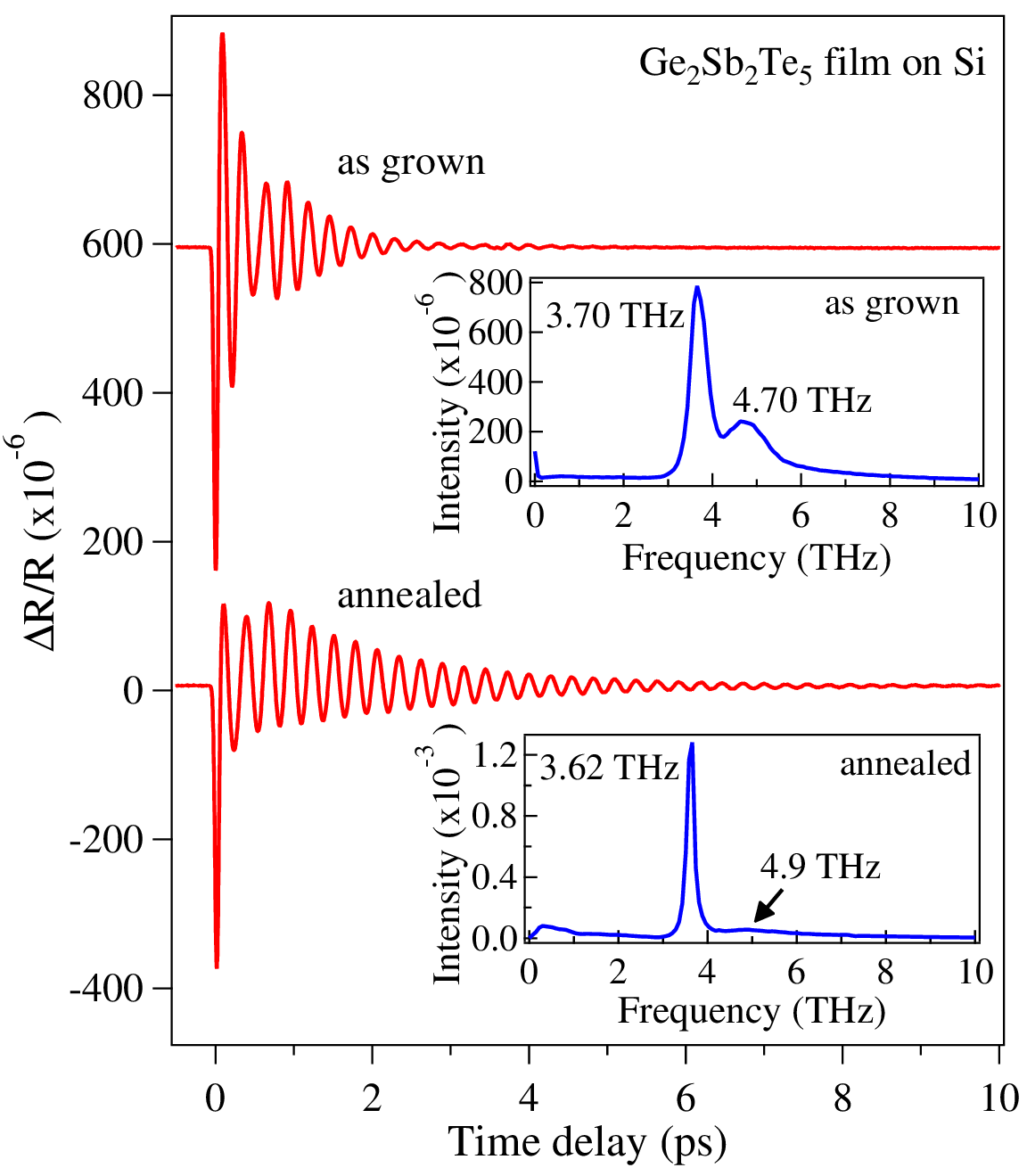}
\caption{(Color online) The TR signal observed in amorphous and crystalline Ge$_{2}$Sb$_{2}$Te$_{5}$ 
films at 295 K. The insets represent FT spectra obtained from the time-domain data. 
}
\label{Fig1}
\end{figure}

Figure 1 shows the time-resolved TR signal ($\Delta R/R$) 
observed in Ge$_{2}$Sb$_{2}$Te$_{5}$ films with amorphous (as-grown) and crystalline 
(annealed) phases at 295 K. After the transient electronic response due to the excitation of 
photo-carriers at $\tau$=0, coherent phonon oscillations with a few picoseconds dephasing 
time appear. Fourier transformed (FT) spectra are obtained from the full scan of the time-domain 
data without any modification as shown in the inset of Fig. 1, in which the two broad peaks 
are observed at 4.70 THz and 3.70 THz in amorphous film, while the sharp peak at 3.62 THz and 
a broad weaker band at $\approx$ 4.9 THz are observed in crystalline film. 
These peaks in the amorphous film can be considered to be the A$_{1}$ optical mode 
due to tetrahedral GeTe$_{4}$ structure for the 3.70 THz peak \cite{Forst00}, and the A$_{1}$ 
optical modes due to disordered Te-Te chains \cite{Forst00,Kolobov06} or the A$_{1}$ optical 
mode due to Sb$_{2}$Te$_{3}$ sublattice as recently been proposed \cite{Andrikopoulos07}, 
 for the 4.70 THz peak. The peaks in the crystalline phase at 3.62 THz and $\approx$ 4.9 THz are 
 almost in 
 agreement with those observed by F\"{o}rst {\it et al.}, although the peak at 2.0 THz is not detected 
 in the present study as observed in the past experiments \cite{Forst00}. 
 The difference in the FT spectra found in the crystalline films would be due to the condition of the 
 sample; we annealed the amorphous GST film at 220 $^{\circ}$C, while the coherent phonon 
 was detected at elevated temperature of 160 $^{\circ}$C in the past work \cite{Forst00}, so the 
 local structure of GST could be slightly different in each case. 

\begin{figure}
\includegraphics[width=8.6cm]{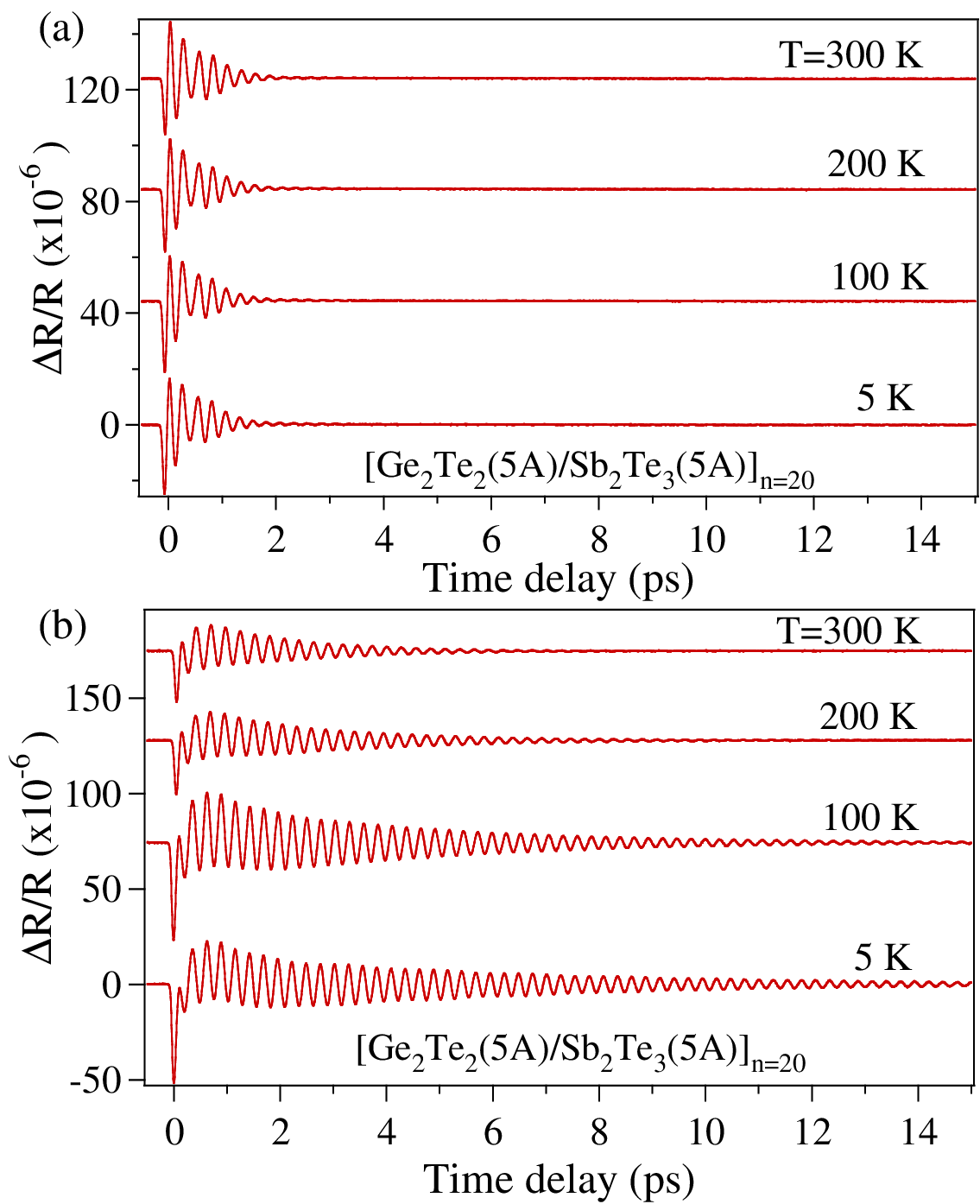}
\caption{(Color online) The TR signal observed in (a) amorphous and in (b) crystalline 
GeTe/Sb$_{2}$Te$_{3}$ SLs at various temperatures. 
}
\label{Fig2}
\end{figure}
Figure 2 compares the time-resolved TR signal observed in 
[Ge$_{2}$Te$_{2}$ (5\AA)/Sb$_{2}$Te$_{3}$ (5\AA)]$_{n=20}$ SL films with (a) amorphous and (b) 
crystalline phases at various temperatures; the total composition corresponds to 
Ge$_{2}$Sb$_{2}$Te$_{5}$ \cite{Tominaga08}. 
In Fig. 2 (a) the coherent phonon oscillations with fast dephasing time of less than $\approx$ 
1.5 ps are observed, and the dephasing time of the coherent phonon seems to not depend on the 
lattice temperature. On the other hand, in Fig. 2 (b) the coherent  oscillation shows longer dephasing 
time than that in Fig. 2 (a) and it is strongly temperature dependent. 

The difference in these two samples is clearer in the corresponding FT spectra, in which 
the two broad peaks are visible at 5.10 THz and 3.78 THz in amorphous film in Fig. 3(a), 
while the dominant 3.68 THz peak is observed in crystalline film in Fig. 3(b) both at 300 K. 
Here, FT spectra are obtained from the full scan of the time-domain data in Fig. 2 without 
any modification. In our experiment, since the observed optical phonons in the amorphous 
films are localized modes,  inhomogeneous damping would partly contribute to the ultrafast 
dynamics of the coherent optical phonons, as reported in glass materials\cite{Guillon05}. 
The inhomogeneous damping would include fluctuation of the environment of the local A$_{1}$  
modes, resulting in the asymmetric line broadening of the FT spectra. The contribution from 
the inhomogeneous damping, however,  
would be negligibly small in the case of GST because the line shapes of the A$_{1}$ modes 
are almost symmetric, which is similar to ion-irradiated Bi, where phonon-vacancy scattering 
dominate the change in the dephasing time. 
The peaks in the amorphous phase can be assigned to the A$_{1}$ optical mode due to 
tetrahedral GeTe$_{4}$ structure for the 3.78 THz peak, and the A$_{1}$ optical modes due 
to amorphous Te-Te chains or the A$_{1}$ optical mode due to Sb$_{2}$Te$_{3}$ sublattice 
for the 5.10 THz peak, as discussed in GST film in Fig. 1 \cite{Forst00,Andrikopoulos07}.  
Note that the peak at 3.68 THz in crystalline phase is slightly lower than that observed in 
amorphous phase (3.78 THz). The peak shift from 3.78 THz (amorphous) to 3.68 THz 
(crystalline) could possibly be due to the change in the local structure of GeTe$_{4}$ into 
GeTe$_{6}$ \cite{Kolobov04}.
It is to be noted that in the crystalline SL a broad peak at $\approx$ 5.0 THz and two sharp 
peaks at 4.37 THz and $\approx$ 7.5 THz are detected, as shown in the inset of Fig. 3 (b). 
The peak at 5.0 THz would be due to the residual of the Te-Te chains or of the optical mode 
from Sb$_{2}$Te$_{3}$ sublattice. The peak at 4.37 THz is not unknown at present, but may 
be a confined A$_{1}$ mode from the Sb$_{2}$Te$_{3}$ layer. The peak at 7.5 THz is possibly 
due to Ge-Ge stretching vibrations as observed in the Raman study \cite{Andrikopoulos07}. 

\begin{figure}
\includegraphics[width=8.6cm]{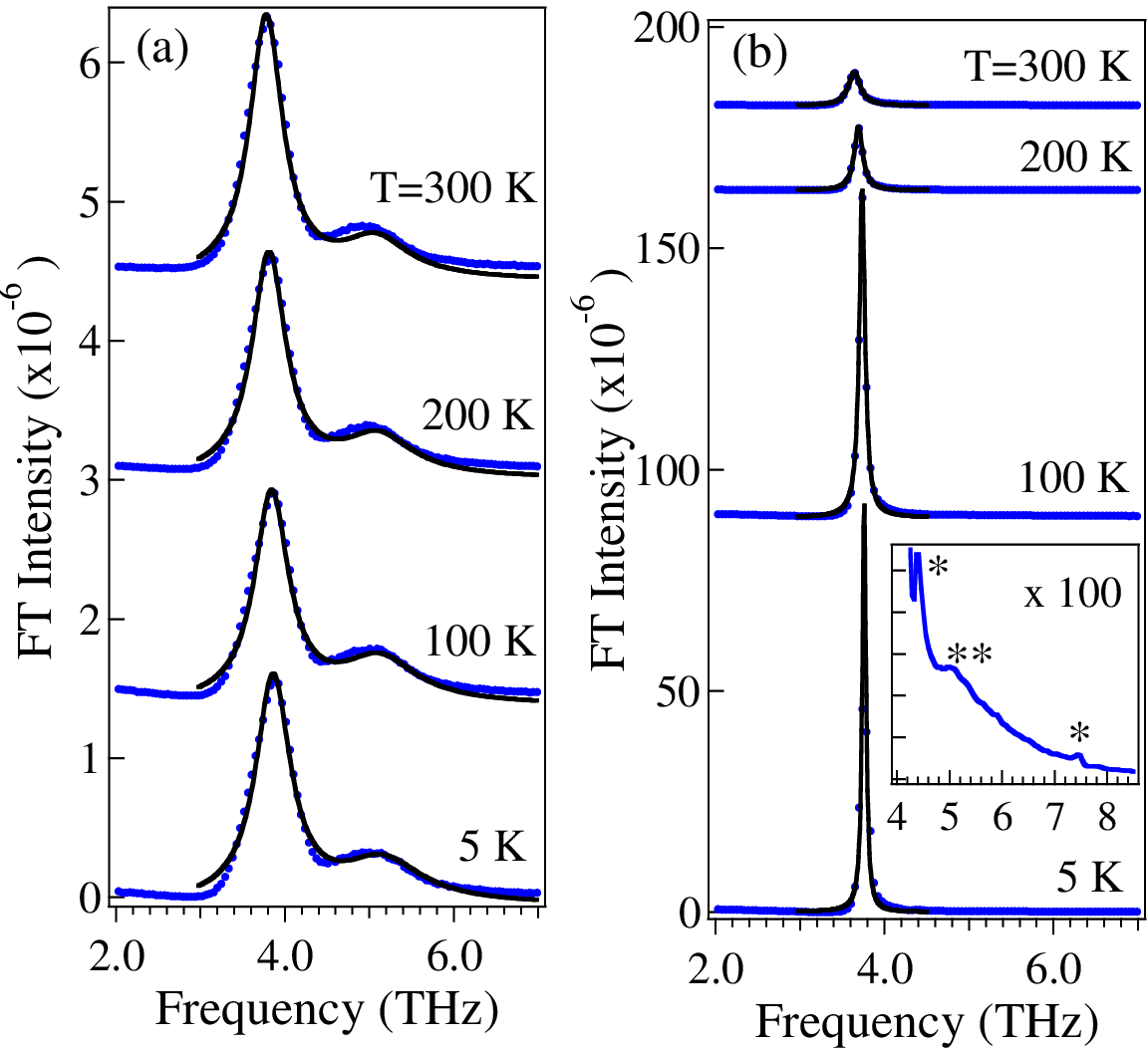}
\caption{(Color online) FT spectra obtained from the time-domain data in Fig. 2; (a) amorphous 
and (b) crystalline GeTe/Sb$_{2}$Te$_{3}$ SLs at various temperatures. The solid lines are 
the fit to the data with Lorentz functions. The inset in (b) represents magnified FT spectra at 
5 K, in which two sharp peaks $(*)$ and a broad peak $(**)$ are detected.  
}
\label{Fig3}
\end{figure}
The coherent optical phonons in GeTe/Sb$_{2}$Te$_{3}$ SLs with amorphous 
phase exhibit significant shift of their frequency relative to the GST film at $\approx$ 300 K. 
The frequency of the A$_{1}$ optical mode at 3.78 THz observed in amorphous SLs was 
originally at 3.70 THz, and that at 
5.10 THz was 4.70 THz in amorphous GST film. These phonon frequencies 
observed in SLs in amorphous phase are 3 $\sim$ 7 \% higher than those observed 
in the GST films, and cannot be attributed to the confinement of the optical phonon modes 
in each layers since the confinement of the optical phonons usually lower the 
frequency \cite{Cardona}.  A plausible explanation for the frequency shift is the volume expansion 
of the amorphous GST film, which is due to the 
randomly distributed vacancies in the amorphous phase, resulting in the reduction of the 
phonon frequency due to the longer bond lengths (smaller bond strength) \cite{Yu99}. 
In fact, based on the first-principle calculations, the volume expansion of $\approx$ 4\% is 
expected due to the randomly distributed vacancies (or voids) in the amorphous phase of GST \cite{Silva08}; 
in amorphous SLs the structure is layered (ordered) and therefore the randomness of 
vacancies will be slightly compensated. In the crystalline phase, on the other hand, 
the single peak observed at 3.68 THz in crystalline SLs is almost the same position to 
that obtained in cubic GST films (3.62 THz in Fig. 1). This would demonstrate that the 
intrinsic structure of the cubic GST film is already the layered structure of 
$[-(Sb_{2}Te_{3}) \cdot \cdot \cdot(Ge-Te-Te-Ge) \cdot \cdot \cdot]_{n}$ \cite{Tominaga08}. 

\begin{figure}
\includegraphics[width=8.6cm]{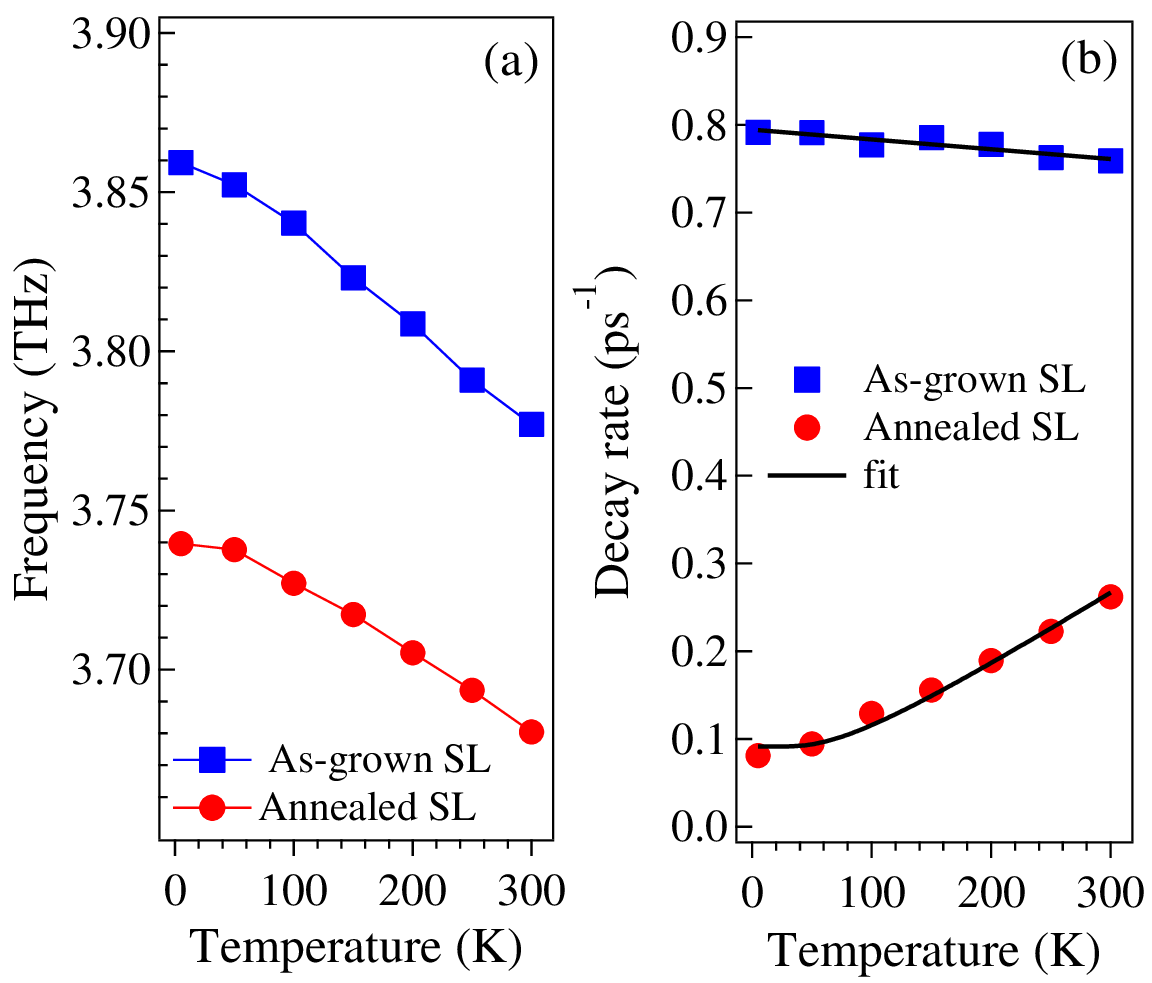} 
\caption{(Color online) The frequency (a) and the decay rate (b) of the coherent A$_{1}$ 
mode, which is localized in GeTe layer, in amorphous and crystalline GeTe/Sb$_{2}$Te$_{3}$ SLs 
as the function of the 
lattice temperatures. In (b) the solid lines are the fit to the data with a linear function for 
the as-grown SL and the anharmonic decay model [Eq. (1)] for the annealed SL. }
\label{Fig4}
\end{figure}

Figure 4 shows the frequency and the decay rate of the coherent A$_{1}$ mode (GeTe$_{4}$ 
or GeTe$_{6}$ modes) 
as a function of the lattice temperature. The decay rate of the crystalline phase 
increases with increasing the temperature, while that in the amorphous phase is almost 
constant when the temperature is varied. The behavior of the decay rate in the crystalline 
phase is well explained by the anharmonic decay model \cite{Vallee94}, in which the optical 
phonon decays into the two acoustic phonons under the conservation of energy and the 
momentum (See Fig. 5); the acoustic phonons with half the 
frequency of the optical mode ($\hbar\Omega_{0}/2$) and with opposite
wavevectors \cite{Vallee94,Hase98}, 
\begin{equation}
\Gamma=\Gamma_{0}\Bigl[1+\frac{2}{exp(\frac{\hbar\Omega_{0}/2}{k_{B}T}
)-1}\Bigr].
\end{equation}
Here $\Gamma_{0}$ is the effective anharmonicity as the fitting parameter and
$k_{B}$ the Boltzmann constant. $\Gamma_{0}$ is determined to be $\approx$ 0.09
ps$^{-1}$. The good agreement of the time domain data with the anharmonic
decay model indicates that the damping of the coherent A$_{1}$ mode in crystalline 
GeTe/Sb$_{2}$Te$_{3}$ SL is due to anharmonic phonon-phonon coupling (population 
decay). The damping in the amorphous phase, on the other hands, does not depend on 
the temperature \cite{Note1} and therefore would be dominated by phonon-defect 
scattering (pure dephasing), whose rate is proportional to the density of lattice defects 
(vacancy) \cite{Hase00,Ishioka01}. 
This supports the conclusion that the randomly distributed vacancies (or voids) 
exist in the amorphous phase, while they change into ordered structure, which 
would not effectively scatter coherent optical phonons. 
Note that the anharmonic phonon-phonon coupling would 
partly contribute to the damping of the A$_{1}$ mode also in the amorphous phase, 
however, the almost flat temperature dependence of the decay rate in Fig. 4 strongly 
suggests that the anharmonic phonon decay path is blocked by the randomly distributed 
vacancies (or voids). Although the calculation of the phonon dispersion for the amorphous 
GeTe is not currently available, a possible reason why the anharmonic decay process is 
absent in the amorphous GST SLs is a breakdown in the phonon momentum conservation 
within the Brillouin zone\cite{ChingPrado96,Kitajima97}. 
\begin{figure}
\includegraphics[width=7.0cm]{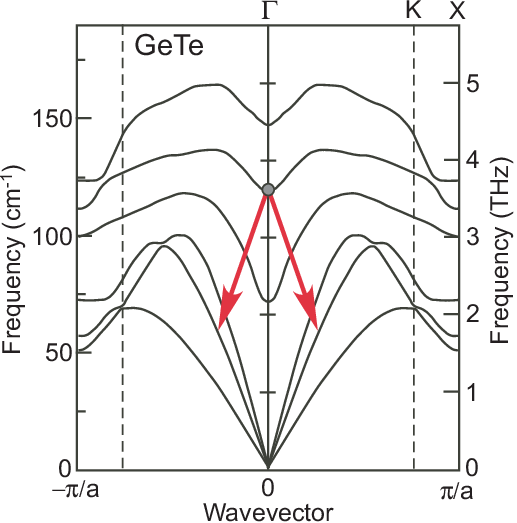} 
\caption{(Color online) Phonon dispersion relations of GeTe. Since the optical phonon 
discussed here is localized in GeTe layer, we assume that the phonon dispersion of the 
GeTe$_{4}$ or GeTe$_{6}$ modes is similar to that of GeTe. The optical mode at $\approx$ 
3.7 THz can then relax into the two underlying acoustic phonons as shown by the arrows. From Ref. [31]. }
\label{Fig5}
\end{figure}

As shown in Fig. 4(a), the frequency of the coherent $A_{1}$ mode decreases as the temperature
increases. This temperature dependence is
qualitatively in good agreement with the anharmonic frequency shift observed
by Raman scattering measurements \cite{Schulz76}. Such a frequency shift due to
the lattice anharmonicity was also observed in III-V semiconductors, which was
reproduced by \textit{ab initio} calculations including various anharmonic
contributions (thermal expansion, third-order, and fourth-order
anharmonicity) \cite{Debernardi00}. 
The difference in the frequency between the amorphous and the crystalline phases 
of $\sim$ 0.1 THz suggests that the local structure of GeTe$_{4}$ in the amorphous phase 
changes into GeTe$_{6}$ in the crystalline phase due to the flip-flop 
structural change in GeTe/Sb$_{2}$Te$_{3}$ SLs \cite{Kolobov04}.

To conclude, we have studied ultrafast dynamics of coherent optical 
phonons in GeTe/Sb$_{2}$Te$_{3}$ SLs  to show the damping of the 
coherent A$_{1}$ mode is temperature dependent in crystalline, while that in the 
amorphous phase does not. 
These facts can be understood in terms of phonon anharmonic decay in the crystalline 
phase, but phonon-defect (vacancy) scattering in the amorphous phase. Thus the existence 
of disordered vacancies (or voids) is evident in amorphous phase, while the vacancies (or voids) in crystalline 
phase are highly ordered. The frequency shift of the A$_{1}$ mode observed in the amorphous 
phase relative to the crystalline phase is suggestive to the local structural change of GeTe$_{4}$ 
into GeTe$_{6}$. The disordering of the vacancies plays dominant role in the volume expansion 
in amorphous GST film, resulting in the frequency red-shift relative to the GeTe/Sb$_{2}$Te$_{3}$ SLs. 
We believe that the present study has uncovered the vivid information on the arrangement of 
the vacancies as well as ultrafast dephasing dynamics of lattice vibrations in 
GeTe/Sb$_{2}$Te$_{3}$ SLs toward the application of laser induced optical switching 
using this unique materials. This method can be applied to all the other Ge-Sb-Te systems 
to understand fundamental lattice dynamics. 

This work was supported in part by KAKENHI-19540329 from MEXT, and "Innovation Research 
Project on Nanoelectronics Materials and Structures - Research and development of superlatticed 
chalcogenide phase-change memory based on new functional structures" from METI, Japan.

\end{document}